\documentclass[%
 reprint,
 amsmath,amssymb,
 aps,
]{revtex4-1}

\usepackage{multirow}
\usepackage{mathrsfs,amsmath}
\usepackage{hyperref}
\usepackage{verbatim}
\usepackage{amsmath}
\usepackage{amsfonts}
\usepackage{amssymb}
\usepackage{amsthm}
\usepackage{bm}
\usepackage{xparse}
\usepackage{graphicx}
\usepackage{xcolor}
\usepackage{textcase}
\usepackage{url}
\usepackage{lipsum}
\usepackage{appendix}
\usepackage{epstopdf}
\usepackage{footnote}
\usepackage{color}
\usepackage[Symbol]{upgreek}

\begin{document}

\preprint{APS/123-QED}

\title{Hybrid topological evolution of multi-singularity vortex beams: Generalized nature for helical-Ince--Gaussian and Hermite--Laguerre--Gaussian modes}% Force line breaks with \\

\author{Yijie Shen}
\email{shenyj15@mails.tsinghua.edu.cn}
\author{Yuan Meng}%
\author{Xing Fu}%
\author{Mali Gong}%

\affiliation{%
	State Key Laboratory of Precision Measurement Technology and Instruments, Center for Photonic and Electronic, Department of Precision Instruments, Tsinghua University, Beijing, 100084, China}%

\date{\today}% It is always \today, today,

\begin{abstract}
A generalized family of scalar structured Gaussian modes including helical-Ince--Gaussian (HIG) and Hermite--Laguerre--Gaussian (HLG) beams is presented with physical insight upon a hybrid topological evolution nature of multi-singularity vortex beams carrying orbital angular momentum (OAM). Considering the physical origins of intrinsic coordinates aberration and the Gouy phase shift, a closed-form expression is derived to characterize the general modes in astigmatic optical systems. Moreover, a graphical representation, Singularities Hybrid Evolution Nature (SHEN) sphere, is proposed to visualize the topological evolution of the multi-singularity beams, accommodating HLG, HIG and other typical subfamilies as characteristic curves on the sphere surface. The salient properties of SHEN sphere for describing the precise singularities splitting phenomena, exotic structured light fields, and Gouy phase shift are illustrated with adequate experimental verifications. 
\end{abstract}

\maketitle

\section{Introduction}
As distinctive structured light fields with phase singularities, optical vortices carrying orbital angular momentum (OAM) have hatched plenty of modern scientific applications in optical tweezers \cite{1,2,2a}, optical communications \cite{3,4}, quantum entanglement \cite{5,6,7} and nonlinear optics \cite{8,9,10}. Besides the classical Laguerre--Gaussian (LG) beams carrying integer OAM with a single phase singularity, the multi-singularity vortex beams carrying fractional OAM were also reported \cite{11,12,13,14,15,16}. The unique characteristics of multi-singularity beams and fractional OAM can be utilized to significantly increase capacity in optical communication systems \cite{4,a1,a1a}, scale multi-particle manipulation technologies in optical tweezers \cite{2,2a,18,19}, drive advanced micro-opto-mechanics \cite{a2}, flexibly shape light beam in 3-dimensional \cite{a3,a4,a5}, and explore novel optical phenomena such as optical vortex knots \cite{20,21,22} and spin-to-orbital conversion \cite{23}. Several theoretical models serving as the eigensolutions of paraxial wave equation (PWE) were established to characterize multi-singularity beams, such as helical-Ince--Gaussian (HIG) modes \cite{31,32}, elliptic LG beams \cite{33}, vortex HG beams \cite{36} and generalized Gaussian beams \cite{37,38}. These models can be attributed to two families: HIG and Hermite--Laguerre--Gaussian (HLG) families, other modes except HIG modes constitute certain subfamilies of HLG family \cite{39,40,41}. HIG and HLG models are capable of describing multi-singularity beams in specific scopes of application. It was theoretically and experimentally verified that the multi-singularity HLG modes are transient states in HG-to-LG mode transforming process via astigmatic mode converter (AMC) \cite{12,14,42,43,44,45,46,47,48}. Also, a large amount of multi-singularity vortex beams that cannot be well interpreted by HLG modes were studied based on HIG modes in terms of Ince-Gaussian (IG) modes \cite{50,51,52,53}. As a family of scalar structured Gaussian modes (SSGMs) that are separable in elliptic coordinates, IG modes were also widely used to characterize special laser beams \cite{31}. Meanwhile, researches showed that IG family has great advantage to characterize the general output modes in laser resonators, which accommodates HG and LG modes \cite{32,50,49a}. Although both HLG and HIG models have their own peculiarities revealing different physical origins for describing multi-singularity modes, the variegated structured Gaussian beams can always show more complexities beyond the scope that classical theoretical modes can describe. Hence, it is significant to investigate generalized theories to illustrate the multi-singularity mode evolution and overcome the limitations of HIG and HLG models.

\begin{figure*}
	\centering
	\includegraphics[width=\linewidth]{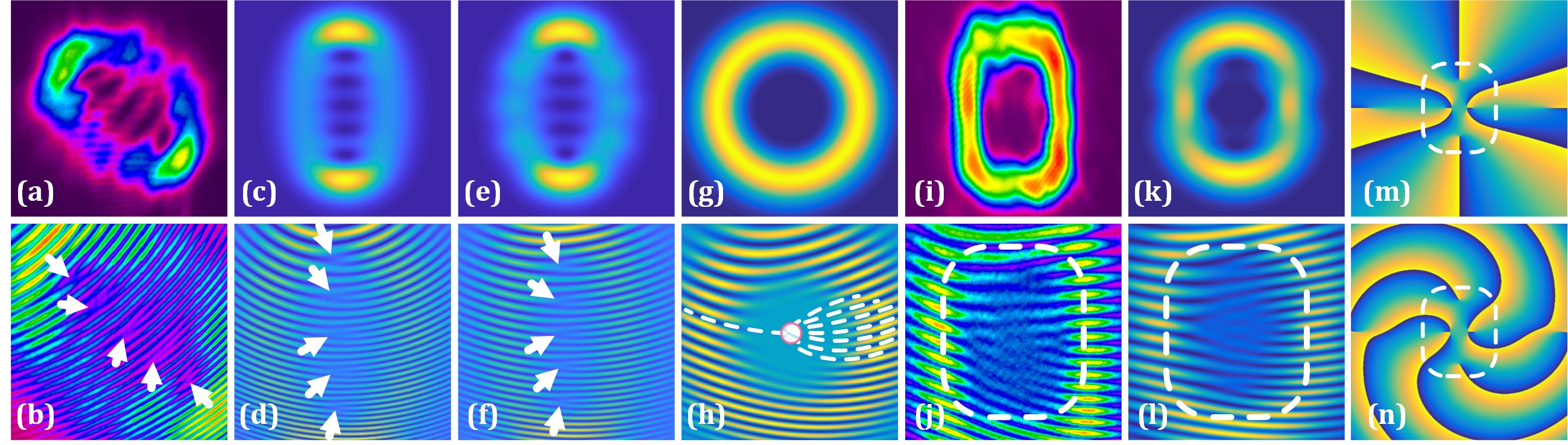}
	\caption{\label{f1} The experimental and theoretical transverse patterns and interferograms: (a,b) the experimental output mode through $\pi/2$ AMC when $\ell=5$ and $\alpha=30^\circ$ and (c,d) the corresponding theoretical interpretations by $\text{HLG}_{0,\ell}\left( \alpha =30^\circ \right)$ mode; (e,f) anothor theoretical interpretations by $\text{HIG}_{\ell,\ell}^{+}\left( \epsilon =4 \right)$ for the linear-multi-singularity mode; (g,h) the theoretical $\text{LG}_{0,\ell}$ mode when $\ell=6$ and $\alpha=45^\circ$; (i,j) the experimental output mode when $\ell=6$ and $\alpha=45^\circ$; (k,l) the theoretical $\text{HIG}_{n,m}^{\bigstar}\left( \epsilon =9 \right)$ mode, and phase distributions (m) at beam waist and (n) at a propagating distance.}
\end{figure*}

In this work, a generalized family of SSGMs including HLG and HIG modes is established to characterize the topological evolution of multi-singularity vortex beams sharing common physical origins in astigmatic optical systems. The description is based on a two-parameter mode family, one parameter represents Gouy phase difference introduced by astigmatism, and the other represents the degree of aberration of the elliptic intrinsic coordinates. Moreover, a novel graphical representation, Singularities Hybrid Evolution Nature (SHEN) sphere, is presented for vividly mapping the evolution of various multi-singularity modes. To verify our model, an abundance of examples are also demonstrated applying SHEN sphere for tailoring various intriguing multi-singularity modes.

\section{Reviews on HLG and HIG modes}
\subsection{Multi-Singularity HLG modes}
A $\pi/2$ AMC comprised of two cylindrical lenses leads to a Gouy phase shift difference of $\pi/2$ at two orthogonal directions of input beam, which can transfer a $\text{HG}_{n,m}$ beam with a $45^\circ$ inclined principal axes into a $\text{LG}_{p,\ell}$ beam, where $p=\text{min}(n,m)$ and $\ell=n-m$, carrying OAM of $\ell\hbar$ per photon, $\ell$ is called topological charge. Basic mode matching conditions are required: 
(a) the cylindrical lenses with focal length of ${f}_{c}$ are separated by a distance of ${d}_{c}={f}_{c}/{\sqrt{2}}$; (b) the input beam has a proper Rayleigh range of ${z}_{R}=(1+{1}/{\sqrt{2}}){f}_{c}$; (c) the beam waist locates at the center of two cylindrical lenses \cite{42,43}. The mode conversion feature is revealed by Beijersbergen's equalities \cite{43}:
\begin{align}
\label{e1}
\text{HG}&_{n,m}\left( \frac{x+y}{\sqrt{2}},\frac{x-y}{\sqrt{2}},z\right)\nonumber\\
&=\sum\limits_{K=0}^{N}{b\left(n,m,K \right) }\cdot{\text{HG}_{N-K,K}}\left(x,y,z\right),
\end{align}
\begin{align}
\label{e2}
\text{LG}&_{p,\pm\ell}\left(x,y,z\right)\nonumber\\&=\sum\limits_{K=0}^{N}{{{\left(\pm\text{i}\right)}^{K}}b\left(n,m,K\right)}\cdot\text{HG}_{N-K,K}\left(x,y,z\right),
\end{align}
where $N=n+m$. Eq.(\ref{e1}) elucidates that a $45^\circ$ diagonal inclined $\text{HG}_{n,m}$ mode can be spaned by the $\text{HG}_{N-K,K}$ modes that constitute $N$-degenerate family with weight coefficients
\begin{equation}
\label{e3}
b\left(n,m,K\right)=\sqrt{\frac{\left( N-K \right)!}{{{2}^{N}}n!m!K!}}\cdot{{\left. \frac{{{\text{d}}^{K}}}{\text{d}{{t}^{K}}}\left[ {{\left( 1-t \right)}^{n}}{{\left( 1+t \right)}^{m}} \right] \right|}_{t=0}},
\end{equation}
then Eq.(\ref{e2}) reveals that a diagonal $\text{HG}_{n,m}$ mode can be converted to $\text{LG}_{p,\pm\ell}$ mode, where the phase shift term $(\pm \text{i})^{K}$ is induced by $\pi/2$ AMC. Assuming an arbitrary inclined angle $\alpha$ of the input $\text{HG}_{n,m}$, the output mode can be expressed by HLG mode having linear splitting singularities (vortices array) with number of $|\ell|$, which is interpolated between HG and LG modes (for $|\ell|>1$) \cite{39,40,41}:
\begin{align}
\label{hlg}
\text{HLG}&_{n,m}\left(\left.x,y,z\right|\alpha  \right)\nonumber\\&=\frac{1}{\sqrt{{{2}^{N-1}}n!m!}}\exp\left( -\frac{{{r}^{2}}}{{{w}^{2}}}\right)\text{HL}_{n,m}\left(\left. \frac{\mathbf{r}}{\sqrt{\pi}w}\right|\alpha\right)\nonumber\\
&\quad\exp\left[ \text{i}kz+\text{i}k\frac{{{r}^{2}}}{2R}-\text{i}\left(m+n+1 \right)\psi\right],
\end{align}
where the Hermite--Laguerre (HL) polynomials
\begin{equation}
\label{hl1}
{\text{HL}\left(\left.\mathbf{r}\right|\alpha  \right)}_{n,m}=\left.\frac{{{\partial}^{m}}}{\partial s_{x}^{m}} \frac{{{\partial }^{n}}}{\partial s_{y}^{n}}\mathcal{G}\left(\left.\mathbf{r},\mathbf{s}\right|\alpha\right)\right|_{\mathbf{s}=\mathbf{0}},
\end{equation}
with generating function
\begin{equation}
\label{hl2}
\mathcal{G}\left(\left.\mathbf{r},\mathbf{s}\right|\alpha\right)=\exp\left[-{{\left({{\mathbf{U}}^{*}}\mathbf{s} \right)}^\mathrm{T}}\left(\mathbf{Us}\right)+2\sqrt{2\pi }{{\left({{\mathbf{U}}^{*}}\mathbf{s} \right)}^\mathrm{T}}\mathbf{r}\right],
\end{equation}
where $\mathbf{r}={{\left(x,y\right)}^\mathrm{T}}={{\left(r\cos \varphi,r\sin\varphi\right)}^\mathrm{T}}$, $\mathbf{s}={{\left( {{s}_{x}},{{s}_{x}}\right)}^\mathrm{T}}$, $\mathbf{U}=\left(\begin{matrix}
\cos\alpha&\text{i}\sin\alpha\\
\text{i}\sin\alpha&\cos\alpha\\
\end{matrix}\right)$, $k={2\pi}/{\lambda}$, $R\left(z\right)={\left(z_{R}^{2}+{{z}^{2}} \right)}/{z}$, the $1/{\rm e}$ beam spot radius $w\left(z \right)=[{2{\left(z_{R}^{2}+{{z}^{2}}\right)}/{\left({{z}_{R}}k \right)}}]^{1/2}$, Gouy phase $\psi\left(z\right)=\arctan\left( {z}/{{{z}_{R}}}\right)$, and $z_R$ is the Rayleigh length. When $\alpha=0$ or $\pi/2$, the $\text{HLG}_{n,m}$ mode is reduced to $\text{HG}_{n,m}$ or $\text{HG}_{m,n}$ mode; when $\alpha = \pi/4$ or $3\pi/4$, the $\text{HLG}_{n,m}$ mode is reduced to $\text{LG}_{p,\pm \ell}$ mode. Fig.\ref{f1}(a,b) shows the measured output mode and their interferogram patterns via a $\pi/2$ AMC when the input was HG$_{0,\ell}$ ($\ell=5$) mode inclined by $30^\circ$ in our experiment (see details in section 5.1), where a linear array of five singularities is illustrated, which are well interpreted by $\text{HLG}_{0,\ell}\left( \alpha ={{30}^\circ} \right)$, as shown in Fig.\ref{f1}(c,d).

Considering that the mode matching conditions (a-c) are not strictly satisfied, i.e. the Gouy phase difference along two orthogonal directions is an arbitrary value $\beta\in(-\pi,\pi]$, while the input is still a $45^\circ$ inclined HG mode, the output mode can be ascribed into the $\beta$-drived HLG family that
\begin{align}
\label{ghlg}
\text{HLG}&_{n,m}^{\bigstar}\left(x,y,z\left|\beta\right. \right)\nonumber\\
&=\sum\limits_{K=0}^{N}{\text{e}^{\text{i}\beta K}b\left(n,m,K\right)}\cdot{\text{HG}_{N-k,k}}\left( x,y,z \right).
\end{align}
In AMC, $\beta$ is related to the $d_c$ and $f_c$ by \cite{43}:
\begin{align}
\beta&=\pm\left(\Delta{{\psi}_{x}}-\Delta{{\psi}_{y}} \right)\nonumber\\
&=\pm2\left(\arctan \sqrt{\frac{1+{{{d}_{c}}}/{{{f}_{c}}}\;}{1-{{{d}_{c}}}/{{{f}_{c}}}\;}}-\arctan \sqrt{\frac{1-{{{d}_{c}}}/{{{f}_{c}}}\;}{1+{{{d}_{c}}}/{{{f}_{c}}}\;}} \right),
\end{align}
where the ``$\pm$'' denotes the interchangeability of $x$, $y$. For $\beta =\pm{\pi}/{2}$, i.e. the case of $\pi/2$ AMC, Eq.(\ref{ghlg}) is reduced to Eq.(\ref{e2}); for $\beta =\pi$ and $\beta=0$, the transformations represent the reversion and invariability respectively. In fact, $\text{HLG}^{\bigstar}$ and $\text{HLG}$ represent the equivalent family but driven by different parameters, sharing the same physical origin of Gouy phase difference \cite{44,45,46}.

\subsection{Multi-Singularity HIG modes}
\begin{figure}[htbp]
	\centering
	\includegraphics[width=\linewidth]{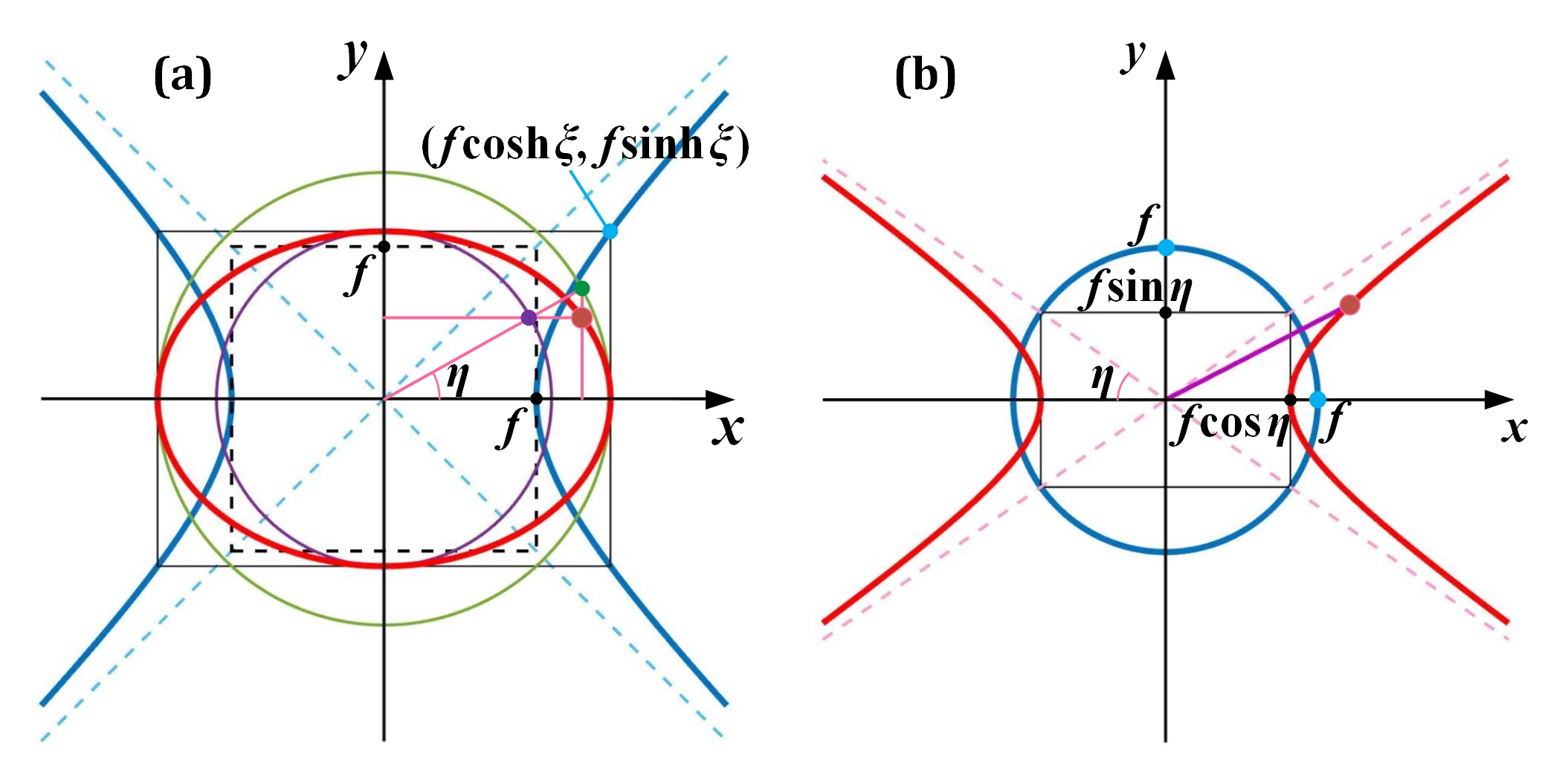}
	\caption{\label{fa5} The intrinsic coordinates system of IG beam can be regarded as a family of (a) ellipse or (b) hyperbola: (a) the ellipse with eccentric angle of $\eta$ and two semi-major axis lengths of $f\cosh\xi$ and $f\sinh\xi$ can cover the whole ${{\mathbb{R}}^{2}}$ plane driven by the point $\left( f\cosh\xi ,f\sinh\xi  \right)$ on a hyperbola; (b) the hyperbola with asymptotic line inclined by $\eta$ angle, semi-real axis lengths of $f\cos\eta$, and semi-imaginary axis lengths $f\sin\eta$ can cover the whole ${{\mathbb{R}}^{2}}$ plane driven by the point $\left( f\cos\eta ,f\sin\eta \right)$ on a circle.}
\end{figure}

The HLG modes cannot completely cover SSGMs. There are other eigensolutions which are separable in elliptic coordinates, e.g. even and odd IG beams \cite{31},
\begin{align}
\label{ige}
\text{IG}_{u,v}^{e}\left(\left.x,y,z \right|\epsilon  \right)=&\frac{{{C}_{\text{IG}}}}{w}C_{u}^{v}\left(\text{i}\xi ,\epsilon\right)C_{u}^{v}\left(\eta,\epsilon\right)\exp\left( -\frac{{{r}^{2}}}{{{w}^{2}}}\right)\nonumber\\
&\exp \left[ \text{i}kz+\text{i}k\frac{{{r}^{2}}}{2R}-\text{i}\left(u+1 \right)\psi\right],
\end{align}
\begin{align}
\label{igo}
\text{IG}_{u,v}^{o}\left(\left.x,y,z \right|\epsilon  \right)=&\frac{{{S}_{\text{IG}}}}{w}S_{u}^{v}\left(\text{i}\xi ,\epsilon\right)S_{u}^{v}\left(\eta,\epsilon\right)\exp\left( -\frac{{{r}^{2}}}{{{w}^{2}}}\right)\nonumber\\
&\exp \left[ \text{i}kz+\text{i}k\frac{{{r}^{2}}}{2R}-\text{i}\left(u+1 \right)\psi\right],
\end{align}
which are also interpolated between HG and LG beams but separable in elliptic coordinates $\left(\xi,\eta\right)$ with ellipticity of $\epsilon$, where ${C}_{\text{IG}}$ and ${S}_{\text{IG}}$ are normalization constants, $C_{u}^{v}\left( \cdot ,\epsilon  \right)$ and $S_{u}^{v}\left( \cdot ,\epsilon  \right)$ are the even and odd Ince polynomials with indices of $(u,v)$. It can be noted that Gouy phase is a general term in a same degenerate state no matter for IG or HLG model, i.e. $\exp \left[-\text{i}\left(u+1\right)\psi\right]=\exp\left[-\text{i}\left(m+n+1\right)\psi\right]=\exp\left[-\text{i}\left(2p+l+1\right)\psi \right]$. $\xi\in\left[0\right.,\left. \infty \right)$ and $\eta\in\left[0\right.,\left.2\pi \right)$ are the radial and angular elliptic variables respectively, which are related to the Cartesian coordinates $\left( x,y \right)$ by
\begin{eqnarray}
\label{ee}
(x,y)=\left[f(z)\cosh\xi\cos\eta,f(z)\sinh\xi\sin\eta\right],
\end{eqnarray}
where $f(z)=w(z)\sqrt{{\epsilon}/{2}}$. Eqs.(\ref{ee}) can be regarded as the parameter equation of ellipse or hyperbola family [schematics are illustrated in Fig.\ref{fa5}(a,b)], which can cover the whole ${{\mathbb{R}}^{2}}$ plane driven by different ellipticity $\epsilon$.

It has been proved that the eigen modes of simple spherical laser cavities are eigenfunctions of two-dimensional quantum harmonic oscillators as the solutions of PWE \cite{60,61,62}, which can be analytically expressed as HG functions with Cartesian symmetry \cite{63}, LG functions with circular symmetry \cite{64}, IG functions with elliptic symmetry \cite{65}, and various distribution functions with other special intrinsic coordinates \cite{66,67}. The intrinsic elliptic coordinates tend to have circular symmetry when the ellipticity parameter $\epsilon \to 0$, and have Cartesian symmetry when $\epsilon \to \infty$ \cite{32,63,64,65}. The recently proposed HIG beams, $\text{HIG}_{u,v}^{\pm }=\text{IG}_{u,v}^{e}\pm \text{i}\cdot \text{IG}_{u,v}^{o}$, have described many multi-singularity beams breaking the limitation of HLG family \cite{32,50,51,52,53}. $\text{HIG}^{\pm}$ model also has the potential to interprete a linear vortices array, such as the multi-singularity mode $\text{HIG}_{\ell,\ell}^{\pm}$ with $\ell=5$, as shown in Fig.\ref{f1}(e,f).

\begin{figure*}
	\centering
	\includegraphics[width=\linewidth]{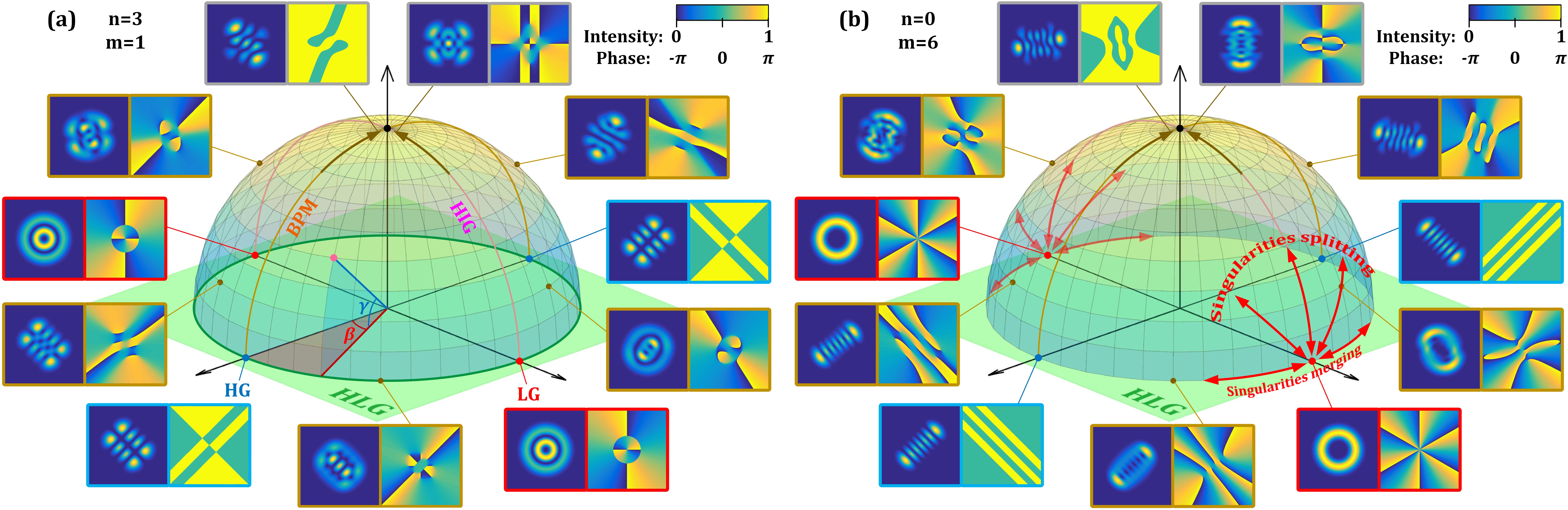}
	\caption{\label{f2} SHEN spheres with orders of (a) $(n,m)=(3,1)$ and (b) $(n,m)=(0,6)$ with represented mode fields (phase) at selected points. The subfamily of HG, LG, HLG, HIG$^{\bigstar}$, BPM modes, and angles $\beta$ and $\gamma$ are marked in figure (a). The tendencies of singularities splitting and singularities merging are marked as double sided arrows in figure (b). See video illustrations of SHEN spheres for more cases: \textbf{Visualization 1} for $(n,m)=(0,1)$, \textbf{Visualization 2} (0,2), \textbf{Visualization 3} (5,0), \textbf{Visualization 4} (6,0), \textbf{Visualization 5} (1,1), \textbf{Visualization 6} (6,6), \textbf{Visualization 7} (1,3), and \textbf{Visualization 8} (6,3).}
\end{figure*}

\subsection{Difficulties of HLG and HIG modes}
HIG$^\pm$ and HLG modes can describe multi-singularity vortex beams or singularities splitting phenomena in AMC systems, however, both of which has limitations (see more evidences in section 5.2). Based on HLG model, doughnut-shape LG$_{0,\ell}$ beam is expected to be obtained via $\pi/2$ AMC for a input $45^\circ$ inclined HG$_{0,\ell}$, as shown in Fig.\ref{f1}(g,h) ($\ell=6$). However, we also observed a square-shape aberrant profile and nonlinear multiple singularities array, see Fig.\ref{f1}(i,j), which is beyond the scope of HLG and HIG$^\pm$ model, even if the AMC was in vicinity of $\beta=\pm\pi/2$ and $\alpha=\pi/2\pm\pi/4$.

In next chapter, we will demonstrate that the elliptical aberration of the input HG modes plays a crucial role in the singularities splitting phenomenon. In other words, the topological evolution of singularities in AMC system is not only dominated by HLG description but also related to an IG aberration principle, coming down to a hybrid topological evolution nature. Moreover, we really observed the corresponding IG aberration evidences in our HG mode oscillator, which is illustrated in section 5.3.

\section{Hybrid Topological Evolution Nature}
\subsection{Hybrid HIG and HLG mode evolution}
For interpreting the exotic singularities splitting phenomenon, we propose a new definition of general HIG mode, which is also defined as the superposition of IG modes but in a more generalized way:
\begin{align}
\label{ghig}
\text{HIG}&_{n,m}^{\bigstar}\left(\left. x,y,z \right|\epsilon\right)\nonumber\\
&=\sum\limits_{K=0}^{N}{{{\left(\pm \text{i}\right)}^{K}}b\left(n,m,K\right)\cdot}\text{IG}_{N,K}^{\bigstar}\left(\left.x,y,z\right|\epsilon\right),
\end{align}
where
\begin{align}
\label{gig}
\text{IG}&_{N,K}^{\bigstar}\left(\left.x,y,z\right|\epsilon  \right)\nonumber\\
&=\left\{\begin{matrix}
{{\left(-\text{i}\right)}^{K}}\text{IG}_{N,N-K}^{e}\left(\left. x,y,z\right|\epsilon\right),\text{for}{{\left(-1\right)}^{K}}=1  \\
{{\left(-\text{i}\right)}^{K}}\text{IG}_{N,N-K+1}^{o}\left( \left.x,y,z\right|\epsilon\right),\text{for}{{\left(-1\right)}^{K}}\ne 1\\
\end{matrix}\right..\quad
\end{align}
In other words, we replace the HG$_{N-K,K}$ modes by the corresponding IG modes that can be degraded into HG$_{N-K,K}$ modes in Eq.(\ref{e2}). When $\epsilon\to\infty$, the $\text{IG}_{N,K}^{\bigstar}$ mode is degraded into HG$_{N-K,K}$ mode, and HIG$^{\bigstar}$ is degraded into LG$_{p,\ell}$. Rather than HLG [Fig.\ref{f1}(g,h)] and $\text{HIG}^{\pm}$ [Fig.\ref{f1}(k,l)] models, the generalized HIG$^{\bigstar}$ model can successfully elucidate the exotic multi-singularity mode shown in Fig.\ref{f1}(i,j), where the precise fringe tracks induced by multiple singularities are highly consistent between experiment and theory. Fig.\ref{f1}(m,n) depicts the simulated phase at beam waist and at a propagating distance respectively, where the morphology of phase vortex solitons is clearly illustrated. It is noted that HLG is included as a subfamily in HIG$^{\bigstar}$ family, therefore, HIG$^{\bigstar}$ description is more felicitous for the actual singularities evolution details in AMC experiment.

Further considering the hybrid properties of HIG$^{\bigstar}$ and HLG modes, a more generalized model is established to elucidate the singularities hybrid evolution in AMC, based on the facts: (a) one usually cannot precisely control an exact phase difference of $\pi/2$ in AMC but an arbitrary $\beta\in\left(-\pi\right.,\left.\pi\right]$; (b) the ellipticity can be evaluated by a parameter $\gamma\in\left[0,\pi/2\right]$ as $\epsilon =2/{\tan^2\gamma}$ and the intrinsic
elliptic coordinates interpolate between Cartesian coordinates and circular coordinates in a harmonic way: 
\begin{equation}
(x,y)=\left(w\cot\gamma\cosh\xi \cos\eta,w\cot\gamma\sinh\xi\sin\eta\right)
\end{equation}
Thus, a more generalized family of SSGMs called as SHEN
family can be established as:
\begin{align}
\label{shen}
&{{\text{SHEN}}_{n,m}}\left(\left.x,y,z\right|\beta,\gamma\right)=\nonumber\\
&\sum\limits_{K=0}^{N}{\text{e}^{\text{i}\beta K}b\left(n,m,K\right)\cdot }\text{IG}_{N,K}^{\bigstar}\left( x,y,z\left|\epsilon =\frac{2}{{{\tan}^{2}}\gamma}\right.\right),\quad
\end{align}
SHEN family is reduced to HIG$^{\bigstar}$ family when $\beta =\pm {\pi }/{2}$, reduced to HLG when $\gamma =0$, to HG when $\left( \beta ,\gamma  \right)=\left( 0,0 \right)$ or $\left( \pi ,0 \right)$, and to LG when $\left( \beta ,\gamma  \right)=\left( \pm {\pi }/{2},0 \right)$ respectively.

\begin{figure*}
	\centering
	\includegraphics[width=\linewidth]{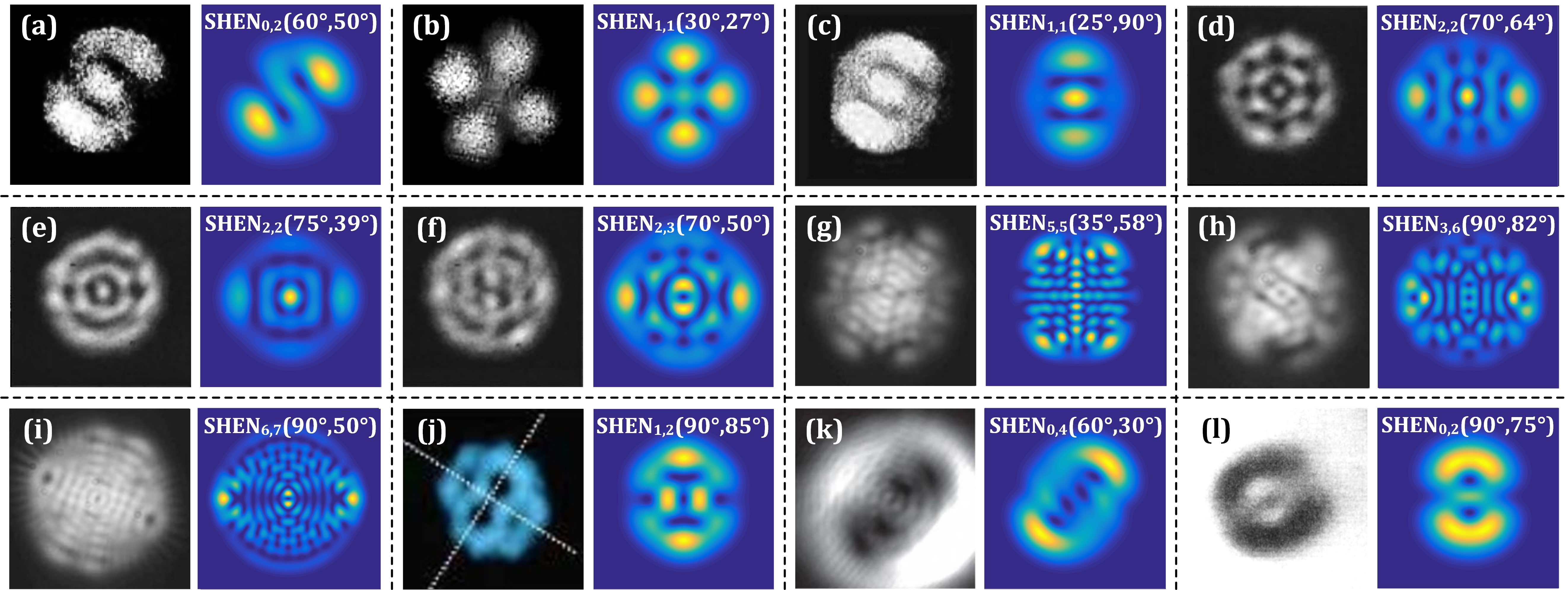}
	\caption{\label{f3} The previously reported experimental modes (left) and the new interpretation by SHEN sphere (right): (a-c) the special modes in Ref.\cite{69}, (d-f) Ref.\cite{70}, (g-i) Ref.\cite{71}, (j) Ref.\cite{72}, (k) Ref.\cite{73}, (l) Ref.\cite{74}.}
\end{figure*}

\subsection{SHEN sphere}
Hereinafter, we present a graphical representation to illustrate the topological evolution of the multi-singularity beams using SHEN sphere. $\text{SHE}{{\text{N}}_{n,m}}\left( \left. x,y,z \right|\beta ,\gamma  \right)$ mode is represented by the point with longitude of $\beta$ and latitude of $\gamma$ on the hemisphere surface. For instance, Figs.\ref{f2}(a,b) illustrates the SHEN spheres with indices of $(n,m)=(3,1)$ and $(n,m)=(0,6)$. Basic properties are noted as follow:
\begin{itemize}	
	\item
	The equator represents the family of HLG$_{n,m}$ modes.
	
	\item
	The $\pm90^\circ$ meridian represents the HIG$^*_{n,m}$ family.
	
	\item
	The prime meridian represents the modes with binary phase distributions, while the binary phase modes (BPM) also have widespread applications such as light field modulation \cite{68}.
	
	\item
	Separated by the prime meridian, the eastern and western quarter-spheres represent two sets of modes which have inversion transformation (chirality) at antipodal points, for both phase and intensity distributions .
	
	\item
	The two points of intersection between the prime meridian and the equator represent the two inversed HG$_{n,m}$ modes.
	
	\item
	The westernmost point and the easternmost point represent two LG$_{p,\pm\ell}$ modes with opposite topological charges, which are the doughnut-shape vortex beams with a single singularity.
	
	\item
	The further away the mode is from the LG points, the more multi-splitting tendencies its singularities possess; while the nearer the mode is to the LG points, the more tendencies of merging into a single singularty they possess.
	
\end{itemize}

We note that there is a singular property on the north pole point of SHEN sphere: the represented mode has non-uniqueness but depends on the longitude as well (just like that you cannot tell a time on the north pole of earth). This property is in contrast to the Poincar\'e sphere\cite{55}, the polarization states described by which are not singular on the south and north poles. We stress that the functions of SHEN sphere and Poincar\'e sphere are different in that SHEN sphere characterizes the scalar fields structure and not polarization. With SHEN sphere, it is clear to explain the aforementioned nonlinear singularities array where an IG aberration of input beam leads to a SHEN mode away from the equator. More examples are shown in supplementary multimedia and can be found for detailed demonstration of the general evolution. Therefore, the singularities hybrid evolution nature of multi-singularity vortex beams can be clearly illustrated by SHEN sphere.

\subsection{Advantages of SHEN sphere}
SHEN sphere possesses great potential for interpreting multi-singularity SSGMs in more optical systems associated with general intrinsic symmetry and astigmatism effects besides the AMC systems. For instance, researchers have reported various exotic multi-singularity vortex beams in microchip lasers and astigmatic optical systems, but lack of theoretical descriptions \cite{69,70,71,72,73,74}. The microchip lasers resonator can be seen as a generalized astigmatic optical system, where the self-reproductive modes are solved by PWE under certain intrinsic coordinates while the gain and thermal effects lead to astigmatism effects \cite{69,70,71,72}. Fig.\ref{f3}(a,b,c) shows the exotic S-shape, X-shape, and $\uptheta$-shape modes that were reported in a Nd:YAG Q-switched microchip laser \cite{69}. The authors provided the superposed expressions to describe them: (a) $\sqrt{0.4}\cdot {\text{e}}^{\text{i}{\pi}/{3}}\cdot{{\text{HG}}_{0,2}}+\sqrt{0.15}\cdot{{\text{HG}}_{2,0}}\text{+}\sqrt{0.45}\cdot {{\text{HG}}_{1,1}}$, (b) $\sqrt{0.15}\cdot{{\text{HG}}_{0,2}}+\sqrt{0.85}\cdot {{\text{e}}^{\text{i}{\pi}/{3}}}\cdot{{\text{HG}}_{1,1}}$, (c) $\sqrt{0.6}\cdot{{\text{HG}}_{0,2}}\text{+}\sqrt{0.2}\cdot{{\text{HG}}_{2,0}}+\sqrt{0.2}\cdot{{\text{e}}^{\text{i}{\pi }/{3}\;}}\cdot {{\text{HG}}_{1,1}}$. These modes can be interpreted in simpler ways under SHEN sphere model with (a) SHEN$_{0,2}(\beta=60^\circ,\gamma=50^\circ)$, (b) SHEN$_{1,1}(\beta=30^\circ,\gamma=27^\circ)$, (c) SHEN$_{1,1}(\beta=25^\circ,\gamma=90^\circ)$. Some complex multi-singularity optical vortices that can hardly be expressed by classical modes were previously reported in a microchip laser with a large Fresnel number \cite{70}, as shown in Fig.\ref{f3}(d,e,f), which can be explained by (d) SHEN$_{2,2}(\beta=70^\circ,\gamma=64^\circ)$, (e) SHEN$_{2,2}(\beta=75^\circ,\gamma=39^\circ)$, (f) SHEN$_{2,2}(\beta=70^\circ,\gamma=50^\circ)$. Also, more complex mode structures were obtained in a microchip self-Q-switched laser \cite{71}, as shown in Fig.\ref{f3}(g,h,i), which were described with a numerical model by authors, while the SHEN sphere can analytically interpret them by (g) SHEN$_{5,5}(\beta=35^\circ,\gamma=58^\circ)$, (h) SHEN$_{3,6}(\beta=90^\circ,\gamma=82^\circ)$, (i) SHEN$_{6,7}(\beta=90^\circ,\gamma=50^\circ)$. Moreover, the mode shown in Fig.\ref{f3}(j) reported in a single-frequency microchip lasers \cite{72} can be interpreted by SHEN$_{1,2}(\beta=90^\circ, \gamma=85^\circ)$; the modes shown in Fig.\ref{f3}(k,l) demonstrated in astigmatic optical systems \cite{73,74} can be interpreted by SHEN$_{0,4}(\beta=60^\circ,\gamma=30^\circ)$ and SHEN$_{0,2}(\beta=90^\circ, \gamma=75^\circ)$. Therefore, SHEN sphere provides a simple and powerful way to explore intriguing features of various structured vortex array beams in singular optics.

\section{Discussion}
Taking advantages of the visualized illustrations and demonstration of clearer physical origins, SHEN sphere is expected to be widely utilized for tailoring structured light fields with multiple singularities matching the actual requirments in these vigorous scientific areas. Required arbitrary topological charges and distribution of multiple phase singularities can be found by scanning through the SHEN sphere of corresponding orders, and can be obtained by the realization of corresponding Gouy phase difference of $\beta$ and ellipticity of $\epsilon$. Therefore, SHEN sphere is more than a mathematical peculiarity, and is readily adapted as a powerful design tool for tailoring exotic structured light beams.

To sum up, a generalized family of SSGMs and a graphical representation are proposed for expressing the topological evolution of multi-singularity vortex beams. Comprehensive phenomena of multi-splitting singularities can be precisely characterized using our model. The SHEN modes are capable of interpreting multifarious SSGMs in general optical systems, naturally accounting for the classical families of HIG and HLG. The SHEN sphere opens a door for tailoring a large variety of structured beams with multiple singularities. Bearing in mind that several typical mode families such as HLG, HIG, BPM are respectively one-dimensional characteristic curves on the two-dimensional surface of SHEN sphere, the presented generalized theory significantly deepens the understanding of structured light fields. 

\section{Methods}
\subsection{Materials and instruments}

\begin{figure}
	\centering
	\includegraphics[width=\linewidth]{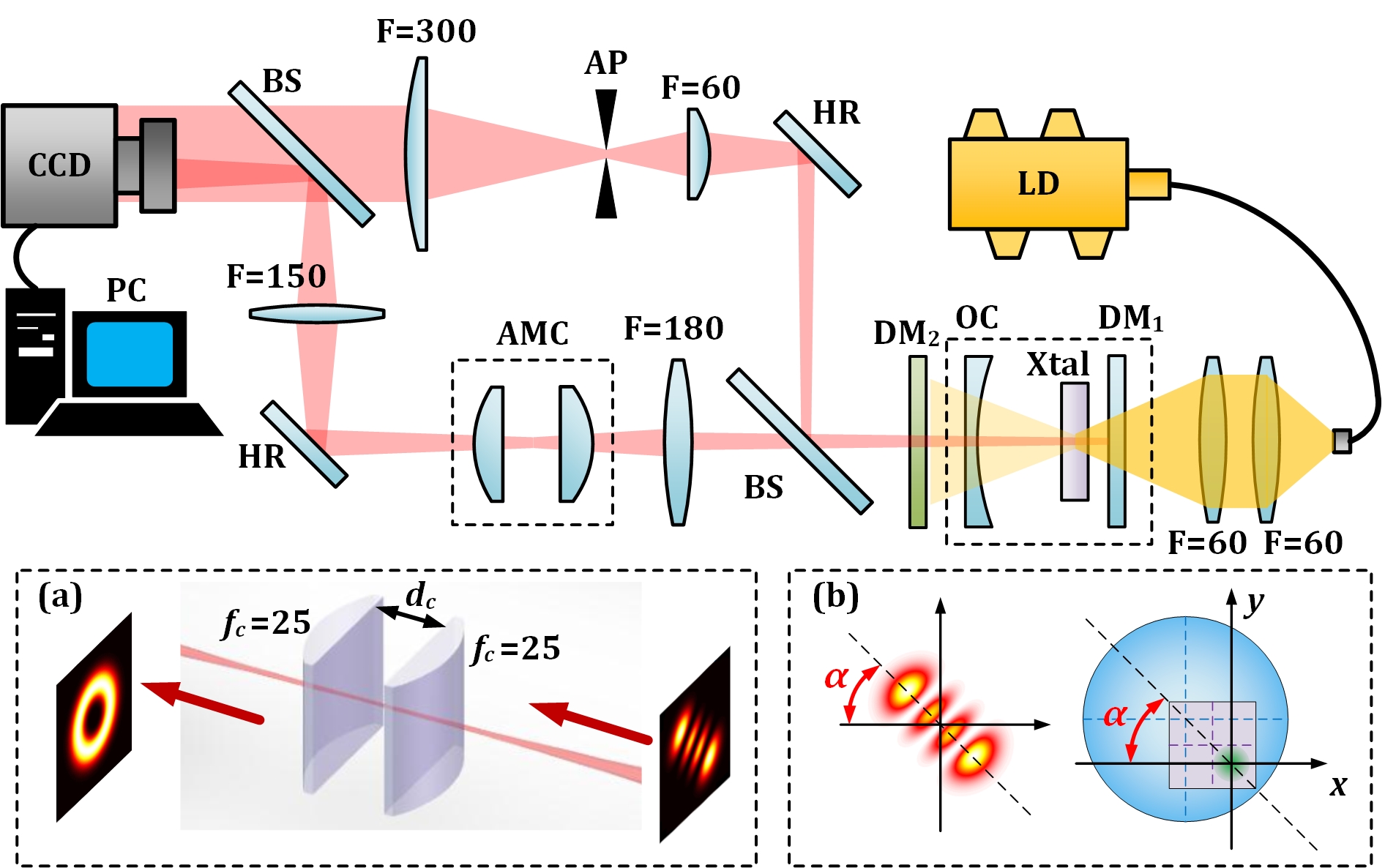}
	\caption{\label{fa1} Schematic of the experimental setup: LD, laser diode; Xtal: crystal; DM, dichroic mirror; OC, output coupler; BS, beam splitter; HR, high-reflective mirror; AP, aperture; AMC, astigmatic mode converter; CCD, charge coupled device; PC, personal computer. Inserts: (a) the details of the AMC with two cylindrical lenses; (b) the methods to control the inclined angle of the output HG mode; units: mm.}
\end{figure}

\begin{figure*}
	\centering
	\includegraphics[width=\linewidth]{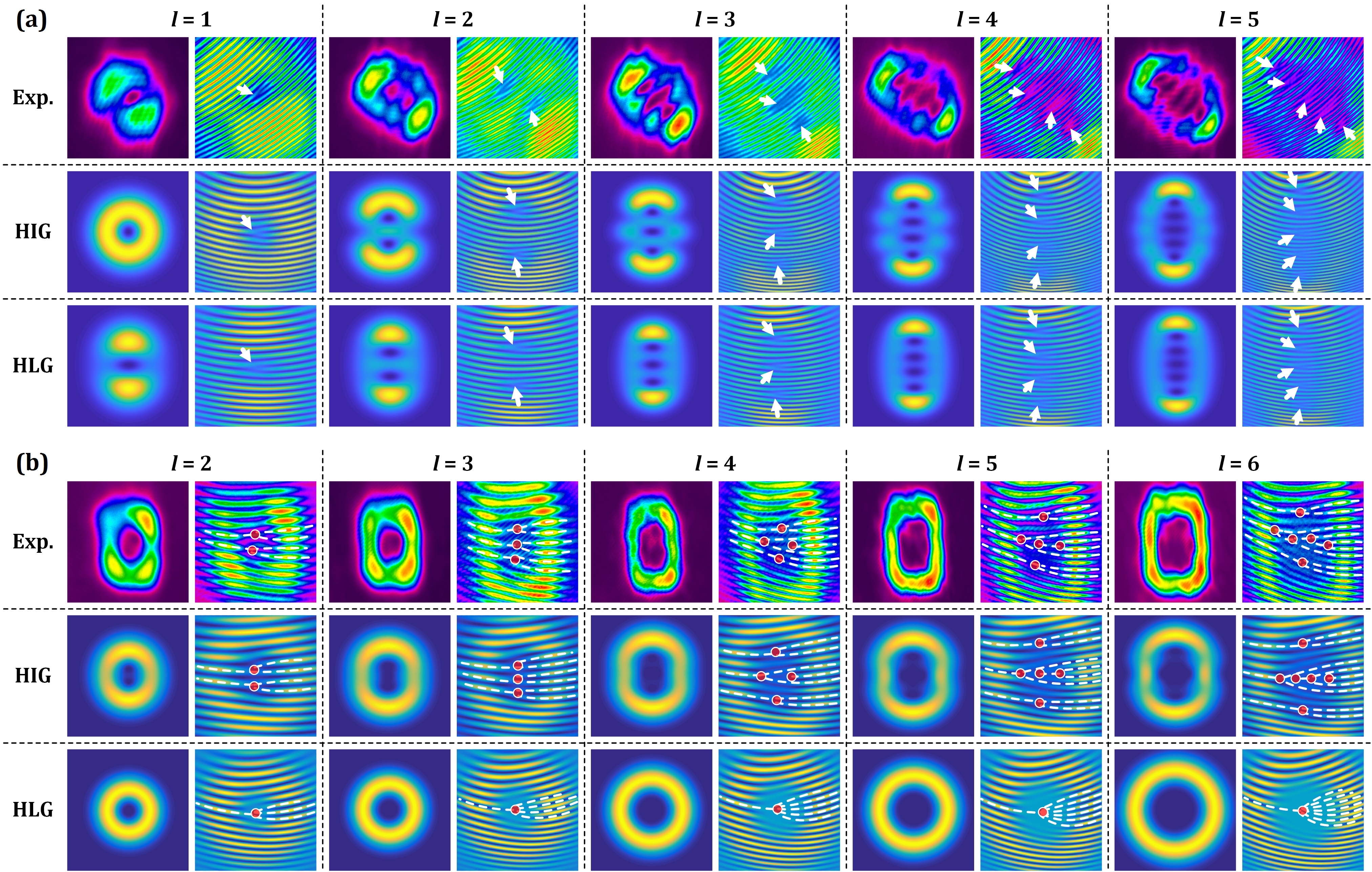}
	\caption{\label{fa3} (a) The first row: the experimental intensity pattern and interferograms for the output mode through AMC with various topological charge $\ell = 1, 2, \cdots , 5$ when inclined angle $\alpha = 30^\circ$; the second row: the corresponding theoretical results of $\text{HIG}_{\ell,\ell}^{+}\left( \epsilon =4 \right)$; the third row: the corresponding theoretical results of $\text{HL}{{\text{G}}_{0,\ell}}\left( \alpha ={{30}^{\text{o}}} \right)$. The positions of singularities are marked with by bold arrows in every interferograms. (b) The first row: the experimental results with various topological charge $\ell = 2, 3, \cdots , 6$ when inclined angle $\alpha = 45^\circ$; the second row: the theoretical results of $\text{HIG}_{0,\ell}^{\bigstar}$; the third row: the theoretical results of $\text{HL}{{\text{G}}_{0,\ell}}\left( \alpha ={{45}^{\text{o}}} \right)$. The positions of singularities are marked with by bold dots and the splitting tracks are marked by dash lines in every interferograms.}
\end{figure*}
%\begin{figure*}[htbp]
%	\centering
%	\includegraphics[width=\linewidth]{Fa4}
%	\caption{\label{fa4} The first row: the experimental intensity pattern and interferograms for the output mode through AMC with various topological charge $\ell = 2, 3, \cdots , 6$ when inclined angle $\alpha = 45^\circ$; the second row: the corresponding theoretical results of $\text{HIG}_{0,\ell}^{\bigstar}$; the third row: the corresponding theoretical results of $\text{HL}{{\text{G}}_{0,\ell}}\left( \alpha ={{45}^{\text{o}}} \right)$. The positions of singularities are marked with by bold dots and the splitting tracks are marked by dash lines in every interferograms.}
%\end{figure*}
The experimental setup, as shown in Fig.\ref{fa1}, includes two main parts: a diode-pumped solid-state laser oscillator for generating HG or IG mode; a Mach-Zehnder interferometer for generating orbital angular momentum (OAM) [via a astigmatic mode converter (AMC) in one arm, see Fig. \ref{fa1}(a)] and observing the phase singularities. 

A 976 nm fiber-coupled laser diode (Hans TCS, core: 105 $\upmu$m, NA: 0.22, highest power: 110 W) was used to pump a Yb:CALGO crystal (Altechna, 2$\times$2$\times$4 mm$^3$, a-cut, 5 at.\%-doped) which was conductively cooling at 18$^\circ$C. The pump waist radius is about 200 $\upmu$m focused by a set of coupling lens including two identical convex lenses with focal length of 60 mm. The laser was generated by a plano-concave resonator with a concave output coupling mirror (OC, radius of curvature of 300 mm) and a flat dichroic mirror [DM$_1$, high-reflection (HR) coated at 1040-1080 nm and antireflective (AR) coated at 976 nm]. The cavity length is 16 cm and the distance between the crystal and DM$_1$ is 2 cm. We used the dual-off-axis method to generate high-order HG or IG modes: the mode order can be controlled by the off-axis distance and pump power, and the inclined angle of mode principal axes can be controlled by the off-axis direction \cite{m1}, as illustrated in Fig. \ref{f1}(b). The DM$_2$ (AR coated at 1040-1080 nm and HR coated at 976 nm) was used to filter residual pump light.
For the interferometer, the two arms were formed by two beam splitters (BS$_1$ and BS$_2$, 45$^\circ$ incidence, T:R=1:9) and two 45$^\circ$ HR mirrors (HR$_1$ and HR$_2$). In one arm, a telescope system was used to obtain near-plane wave as reference light, which includes two convex lenses with focal length of 60 mm and 300 mm respectively and an aperture. In another arm, the laser was focused into AMC by a convex lens with focal length of 180 mm for generating multi-singularity beam. The AMC was composed of two identical convex-plane cylindrical lenses with focal length of $f_c = 25$ mm separated by $d_c = 35.4$ mm ($\sqrt{2}{{f}_{c}}$). The multi-singularity beam was captured by a CCD camera (Spiricon, M2-200s) after being focused by a convex lens with focal length of 150 mm for observing the interference pattern. 

\subsection{Generation of multi-singularity modes}
The multi-singularity beam is generated via the AMC and the input beam is the HG$_{0,m}$ beam generated from the dual-off-axis resonator, which inevitably has IG aberration, as demonstrated in the next section. The corresponding perfect vortex beam converted by $\pi/2$ converter should have the topological charge $\ell=m$. However, perfect $\pi/2$ converter is difficult to obtain because there is usually a Rayleigh range mismatch. The first row of Fig.\ref{fa3}(a) shows the intensity profiles and interferograms for the case with inclined angle of $\alpha = 30^\circ$ and $\ell$ from 1 to 5. The HIG and HLG mode can describe these modes with linear splitting singularities array, i.e. $\text{HIG}_{\ell,\ell}^{+}\left( \epsilon =4 \right)$ and ${{\text{HLG}}_{0,\ell}}\left( \alpha ={{30}^{\text{o}}} \right)$, the simulated results of which (intensity profiles and interferograms) are shown in the second and third row of Fig.\ref{fa3}(a) respectively. The HIG and HLG modes can describe the actual multi-singularity modes for $\ell = 2, 3, 4, 5$ with a good agreement (the linear distributed singularities are marked by bold arrows). However, for $\ell = 1$, the HIG mode shows a doughnut shape that is deviated from the actual dual-hump profile, while HLG modes maintain a good agreement. For the case with the inclined angle of $\alpha = 45^\circ$, as shown in the first row of Fig.\ref{fa3}(b), a square-shape aberrant profile is obtained rather than a perfect LG vortex mode with a circular doughnut profile [as the third row of Fig.\ref{fa3}(b)], and the splitting singularities with number of $\ell$ can still be observed. The first reason is that the input HG mode has Ince-aberration, and the second one is the mode mismatch for realizing $\pi/2$ converter. In these cases, the limitations of HLG modes are revealed: neither the square-shape aberration nor the splitting singularities phenomenon can be well described. For the sake of establishing more generalized model to describe the actual modes evolution in astigmatic optical systems, the aberration degree of intrinsic coordinates should be considered. Overcoming the limitations of HLG and HIG$^\pm$ modes, the HIG$^{\bigstar}$ modes can give better interpretations for the singularities splitting and mode field aberration phenomenon, as shown in the second row of Fig.\ref{fa3}(b).

\subsection{Eigen mode output in laser resonators}
\begin{figure}
	\centering
	\includegraphics[width=\linewidth]{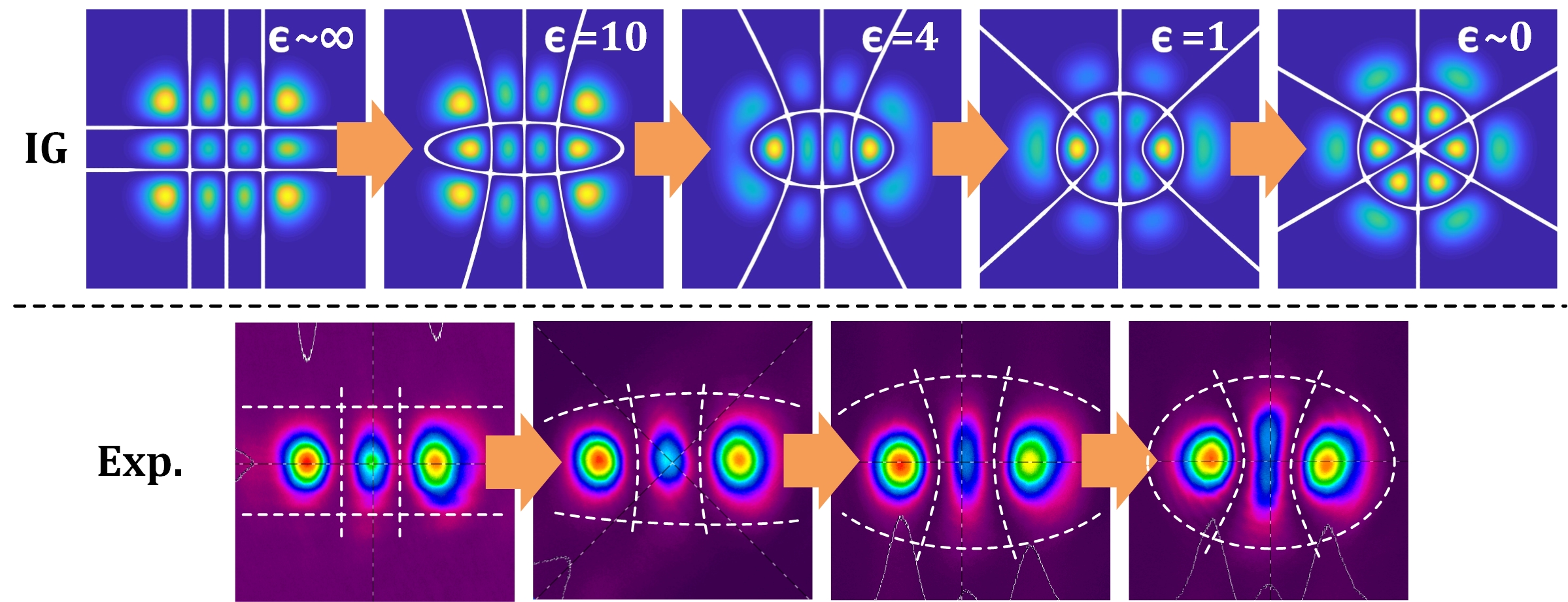}
	\caption{\label{fa2} The first row: the evolution of IG modes with various ellipticity gradually from infinity to zero, i.e. the HG mode is gradually transformed to the even or odd LG mode. The second row: the experimental results that an output HG mode can be transformed to IG mode with a certain ellipticity by slightly adjusting the laser resonator.}
\end{figure}
The output mode in a laser resonator is usually a superior eigen mode in gain competition \cite{m2}, the principle of which has guided many researchers to realize direct HG and LG output from resonator by using various selective pumping methods \cite{m1,m6,m7,m8,m9,m10}. However, the intrinsic symmetry is actually influenced by complex factors such as cavity design, off-axis displacement \cite{m1,m9,m10}, gain and thermal distributions \cite{m11,m12}, as well as astigmatism effects of pump light \cite{m12,49a,50}, all of which can contribute to actual laser output mode. The IG modes, another form of eigenfunctions to the PWE that are separable under elliptic coordinates, can characterize more categories of output modes in laser resonator than HG and LG modes, and the HG and LG modes are just two extreme cases of IG modes \cite{32}, as illustrated in the first row of Fig. \ref{fa2}. The transition from an IG mode to a HG mode occurs when the elliptic coordinates tend to the Cartesian coordinates, i.e., when $\epsilon \to \infty$. The relation between HG and IG modes are given as below:
\begin{align}
&{{\text{HG}}_{n,m}}\left( x,y,z \right)=\nonumber\\
&\left\{ \begin{matrix}
{{\left( -\text{i} \right)}^{m}} \text{IG}_{m+n,n}^{e}\left( \left. x,y,z \right|\epsilon \to \infty  \right),\text{ for }{{\left( -1 \right)}^{m}}=1  \\
{{\left( -\text{i} \right)}^{m}}\text{IG}_{m+n,n+1}^{o}\left( \left. x,y,z \right|\epsilon \to \infty  \right),\text{ for }{{\left( -1 \right)}^{m}}\ne 1  \\
\end{matrix} \right.,
\label{re1}
\end{align}

and the transition from an $\text{IG}_{u,v}^{e,o}$ mode to a $\text{IG}_{p,\ell}^{e,o}$ mode takes place when the elliptic coordinates approach circular cylindrical coordinates, i.e., when $\epsilon \to 0$. The relations between LG and IG modes are given by
\begin{equation}
\text{LG}_{p,\ell}^{e}\left( x,y,z \right)={{\left( -\text{i} \right)}^{m}}\text{IG}_{2p+\ell,\ell}^{e}\left( \left. x,y,z \right|\epsilon \to 0 \right),
\label{re2}
\end{equation}
\begin{equation}
\text{LG}_{p,\ell}^{o}\left( x,y,z \right)={{\left( -\text{i} \right)}^{m+1}}\text{IG}_{2p+\ell,\ell}^{o}\left( \left. x,y,z \right|\epsilon \to 0 \right),
\label{re3}
\end{equation}
where the even and odd LG modes, $\text{LG}_{p,\ell}^{e}$ and $\text{LG}_{p,\ell}^{o}$, are defined by replacing the Hilbert phase term $\exp \left( \text{i}\ell\varphi \right)$ in normalized LG modes with $\cos \left( \ell\varphi  \right)$ and $\sin\left( \ell\varphi \right)$ respectively \cite{32}. The relation of the two kinds of LG modes are $\text{LG}_{p,\ell}^{e}={\left(\text{L}{{\text{G}}_{p,\ell}}+\text{L}{{\text{G}}_{p,-\ell}}\right)}/{2}$ and $\text{LG}_{p,\ell}^{o}={-\text{i}\left( \text{L}{{\text{G}}_{p,\ell}}-\text{L}{{\text{G}}_{p,-\ell}} \right)}/{2}$. The analytical expressions of HG$_{n,m}$ and LG$_{p,\ell}$ modes are given by:
\begin{align}
\text{HG}&_{n,m}\left(x,y,z\right)\nonumber\\&=\frac{C^{\text{HG}}_{n,m}}{w}\exp\left(-\frac{x^2+y^2}{w^2}\right)H_n\left(\frac{\sqrt{2}x}{w}\right)H_m\left(\frac{\sqrt{2}y}{w}\right)\nonumber\\
&\quad\exp\left[ \text{i}kz+\text{i}k\frac{{x^2+y^2}}{2R}-\text{i}\left(m+n+1 \right)\psi\right],
\end{align}
\begin{align}
\text{LG}&_{p,\ell}\left(r,\varphi,z\right)\nonumber\\&=\frac{C^{\text{LG}}_{p,\ell}}{w}\left(\frac{\sqrt{2}r}{w}\right)^{\left|\ell\right|}\exp\left(-\frac{r^2}{w^2}\right)L_p^{\left|\ell\right|}\left(\frac{2r^2}{w}\right)\exp\left(\text{i}\ell\varphi\right)\nonumber\\
&\quad\exp\left[ \text{i}kz+\text{i}k\frac{{r^2}}{2R}-\text{i}\left(2p+\left|\ell\right|+1 \right)\psi\right],
\end{align}
where $(r,\varphi)=[\sqrt{x^2+y^2},\arctan(y/x)]$, $H_n(\cdot)$ is the Hermite polynomial of $n$-th order, $L_p^\ell(\cdot)$ is the generalized Laguerre polynomial with radial and azimuthal indices of $p$ and $\ell$, and the normalization constants for HG and LG modes are $C^{\text{HG}}_{n,m}=\sqrt{{2}/{(\pi m!n!)}}\cdot2^{-N/2}$ and $C^{\text{LG}}_{p,\ell}=\sqrt{{2p!}/{[\pi(p+\left|\ell\right|)!]}}=\sqrt{{2}/{(\pi m!n!)}}\cdot{\text{min}(n,m)!}$.

\section*{Funding}
The National Key Research and Development Program of China (Grant no. 2017YFB1104500); Natural Science Foundation of Beijing Municipality (4172030); Beijing Young Talents Support Project (2017000020124G044).

\end{document}